# Potentials and Weak Potentials in Acyclic and Weakly Acyclic Games

Igal Milchtaich, Department of Economics, Bar-Ilan University
igal.milchtaich@biu.ac.il https://sites.google.com/view/milchtaich

**Abstract** In a number of large, important families of finite games, not only is the set of pure-strategy Nash equilibria nonempty but it is also reachable from any initial strategy profile by some sequence of myopic single-player moves to a better or best-reply strategy. This *weak acyclicity* property is weaker than *acyclicity* of the game, which requires *every* such sequence to reach an equilibrium. For example, all perfect-information extensive-form games are weakly acyclic, but they are generally not acyclic as even sequences of best-improvement steps may cycle. Weak acyclicity is equivalent to acyclicity of some *priority rule*, which is a rule that allows only some improvement moves. It is also equivalent to the existence of a *weak potential*, which unlike a *potential* increases along some rather than every sequence as above. *JEL Classification:* C72.

## Basic Concepts

This paper concerns finite games, with a finite number $n$ of players and a finite strategy set $S_i$ for each player $i$. Correspondingly, "strategy" always means pure strategy. The payoff function of player $i$ is denoted by $u_i$. A *subgame* of a finite game $\Gamma$ is obtained by replacing each strategy set $S_i$ with some subset of $S_i$ and restricting the payoff functions accordingly.[1] In the special case where the strategy sets of some players are reduced to singletons, it is possible to view only the other players as players in the subgame.

The *improvement graph* of a finite game $\Gamma$ is the directed graph that describes the players' profitable unilateral deviations. Its vertices are the strategy profiles in the game, and for every pair of strategy profiles $s$ and $t$, a (directed) edge with head $s$ and tail $t$ exists if and only if there is some player $i$ such that $s_j = t_j$ for all $j \neq i$ (thus, $s = (s_i, t_{-i})$) and

$$u_i(s) > u_i(t). \tag{1}$$

The *best-improvement graph* of $\Gamma$ is the subgraph obtained by adding to (1) the requirement that strategy $s_i$ is a best reply to $s_{-i}$ $(= t_{-i})$, that is, $u_i(s) \geq u_i(s'_i, s_{-i})$ for all $s'_i \in S_i$. Obviously, a strategy profile is a sink of either the improvement or best-improvement graph if and only if it is a (pure-strategy Nash) *equilibrium* in $\Gamma$. In the following, "the graph" of $\Gamma$ and related terms may refer to either graph. This convention enables the simultaneous introduction of two parallel terminologies. However, unless the meaning can be understood from the context, unambiguous use of any term

---

[1] It should be clear from the context whether "subgame" is meant in this sense (Shapley 1964) or in the more familiar one pertaining to extensive-form games (Selten 1975). In particular, the latter holds when the reference is to an extensive-form game $G$ and the former holds when the reference is to the normal, or strategic, form $\Gamma$ of $G$.

requires indicating the graph it refers to, which may be done by prefixing it with either I- or BI-.[2]

A *priority rule* for a finite game $\Gamma$ is any spanning subgraph of the graph of $\Gamma$ that has the same sinks as the latter. Less formally, it is a rule that, for each strategy profile $s$ that is not an equilibrium, (possibly) restricts some players' freedom of choice by allowing them to take only some moves or no moves at all, without making it impossible to leave $s$. For example, the rule may stipulate that certain kinds of moves take precedence over others, so that whenever any of the former is feasible, none of the latter is allowed.[3] One priority rule is *stronger* than another if it is a subgraph of it. The second, *weaker* priority rule allows every move allowed by the first one but not necessarily the other way around.[4] The weakest priority rule, which is the (improvement or best-improvement) graph itself, is referred to as the *trivial* priority rule.

For a given priority rule, a *walk* of length $m$ ($\geq 0$) is any finite sequence $s^0, s^1, \dots, s^m$ of (not necessarily distinct) strategy profiles such that for $l = 1, 2, \dots, m$ there is an edge in the priority rule whose tail and head are $s^{l-1}$ and $s^l$, respectively. A walk is *closed* if $m > 0$ and $s^m = s^0$. It is a *path* if all $m + 1$ strategy profiles are distinct. One walk or path *extends* another if it is obtained from it by appending one or more strategy profiles. A priority rule is *acyclic* if it has no closed walks, and *weakly acyclic* if some stronger priority rule is acyclic.

The game itself is said to be acyclic or weakly acyclic if the trivial priority rule has the same property. A path in the trivial priority rule is also called an *improvement* or *best-*(*reply*) *improvement* path, depending on the graph considered. Correspondingly, alternative terms for the (weak) I- and BI-acyclicity properties of games are the (respectively, *weak*) *finite improvement* and *finite best-*(*reply*) *improvement* properties.1996[5] It is easy to see that the four properties are linearly ordered by the implication relation, as follows:

$$\text{I-acyclicity} \Longrightarrow \text{BI-acyclicity} \Longrightarrow \text{weak BI-acyclicity} \Longrightarrow \text{weak I-acyclicity.}$$

---

[2] A third graph considered in the literature is the *best-reply* graph, which differs from the best-improvement graph in that it describes also moves *between* best-reply strategies. A strategy profile is a sink of the best-reply graph if and only if it is a *strict* equilibrium. The set of edges in the best-improvement graph is the intersection of the sets of edges in the best-reply graph and the improvement graph.

[3] In particular, a priority rule may allow all players to move but restrict their choice of alternative strategies, or it may allow only one player to move at each strategy profile without interfering with the mover's choice of strategy. Priority rules of the first kind are studied by Kukushkin (2004, 2011) in the context of restricted acyclicity. Priority rules of the second kind are called *schedulers* by Apt and Simon (2015). However, the scheduler concept is more general in that the mover's identity may depend on the history of moves and not only the current strategy profile.

[4] These definitions entail reflexivity: every priority rule is both stronger and weaker than itself. The corresponding irreflexive relations may be indicated by the qualifier *strictly*.

[5] Young's (1993) notion of (weak) acyclicity is similar except that it refers to the best-reply graph (see footnote 2) and may therefore be referred to as (respectively, weak) *BR-acyclicity*. BR-acyclicity in particular precludes the existence of *best-response cycles* in the sense of Voorneveld (2000). The latter differ from closed walks in the best-improvement graph in that only one of the changes of strategy is required to be an improvement; the rest may be moves between best-reply strategies.

For some examples of games possessing one or more of these properties see Monderer and Shapley (1996), Milchtaich (1996, 2009), Friedman and Mezzetti (2001), Milchtaich and Winter (2002), Kukushkin et al. (2005), Engelberg and Schapira (2014), and Theorems 2, 3, 4 and 5 below.

A real-valued function $P$ on the set of vertices is a *potential* for a priority rule if it increases along every walk in it, in other words, if for every two strategy profiles $s$ and $t$ that are respectively the head and tail of an edge in the priority rule,

$$P(s) > P(t). \tag{2}$$

A function $P$ is a *weak potential* for a priority rule if it is a potential for some stronger priority rule, equivalently, if the subgraph obtained from the priority rule by eliminating all edges whose head and tail do not satisfy (2) is also a priority rule. A necessary and sufficient condition for this is that every *local minimum point* of $P$, that is, a strategy profile $t$ that is not the tail of any edge in the priority rule whose head $s$ satisfies (2), is an equilibrium.

A potential or weak potential for a *game* means such a function for the trivial priority rule. An alternative term for I-potential for a game, which stresses the distinction between this concept and the related cardinal one of exact potential (Monderer and Shapley 1996), is *generalized ordinal potential*. It is easy to see that the following implications between properties of a function $P$ on the strategy profiles hold:

I-potential ⟹ BI-potential ⟹ weak BI-potential ⟹ weak I-potential.

## Basic Result

The following theorem applies to both the improvement and best-improvement graphs.

**Theorem 1** (Monderer and Shapley 1996, Kukushkin 2004) For a finite game, or more generally a priority rule for such a game, the following properties are equivalent:

(i) acyclicity,
(ii) existence of potential,
(iii) every walk can be extended at most finitely many times (before an equilibrium is reached).

Similarly, the following properties are equivalent:

(i') weak acyclicity,
(ii') existence of weak potential,
(iii') for every strategy profile $s$, some path that starts at $s$ ends at an equilibrium.

*Proof.* For an acyclic priority rule (or, as a special case, an acyclic game), consider for each strategy profile $s$ the length of the longest path that starts at $s$. This number is 0 if and only if $s$ is an equilibrium. Its negative,

$$P(s) = -\max\{m \geq 0 \mid \text{there is a path of length } m \text{ that starts at } s\},$$

defines a potential, as it is easy to see that $P$ increases along any walk. Conversely, for a priority rule that does have a closed walk, a potential clearly does not exist, and the walk can be extended indefinitely by repetition.

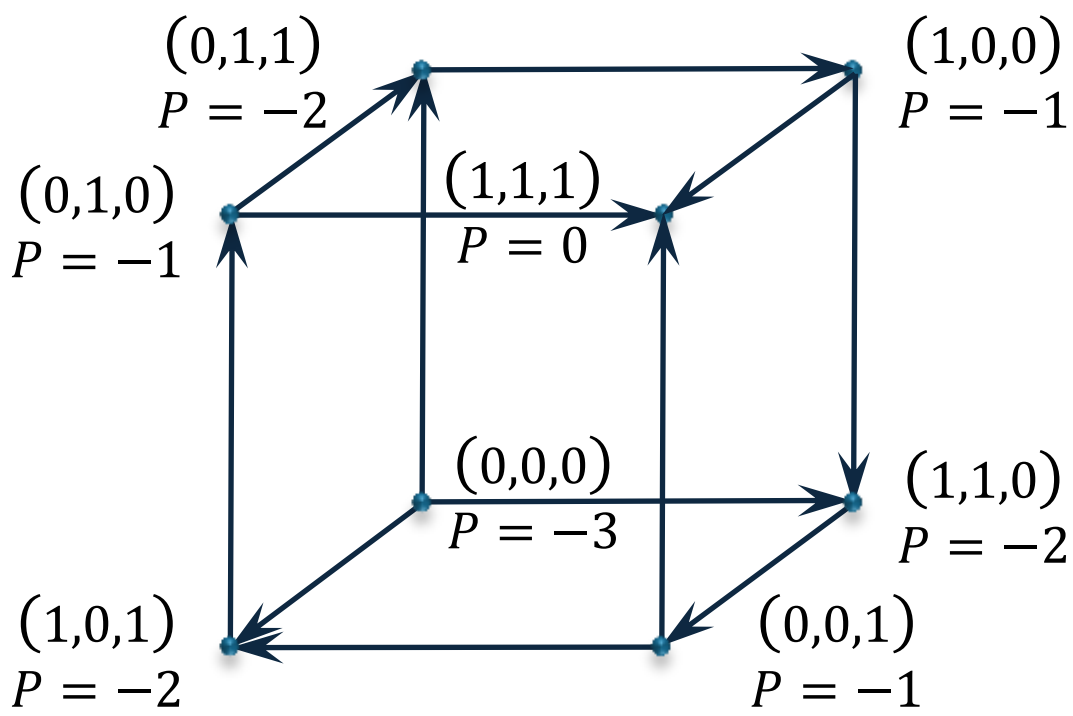

**FIGURE 1 The cube-like improvement (which is also the best-improvement) graph of a finite $2 \times 2 \times 2$ game. The players' strategies are Right and Left for player 1, Up and Down for player 2, and In and Out for player 3. Shown for each strategy profile are the players' payoff vector and the value of a weak potential $P$ defined as in the proof of Theorem 1. This game is (i) weakly acyclic but (ii) not acyclic, (iii) not solvable by iterated elimination of never-best-reply strategies, and (iv) has the property that every subgame (which corresponds to a $k$-dimensional face of the cube, with $0 \leq k \leq 3$) has a unique equilibrium.**

By the first part of the proof, a priority rule is weakly acyclic if and only if some stronger priority rule possesses a potential, in other words, if and only if the priority rule itself possesses a weak potential. In this case, every walk in the stronger priority rule that starts at a given strategy profile $s$ is a path and strategy profiles can be appended to it only finitely many times before an equilibrium is reached, which proves that (iii') holds. Conversely, for a priority rule that satisfies (iii'), consider for each strategy profile $s$ the distance in the priority rule to the closest equilibrium. The negative of this distance,

$$P(s) = -\min\{m \geq 0 \mid \text{there is a path of length } m \text{ that starts at } s \text{ and ends in an equilibrium}\},$$

defines a function $P$ on the strategy profiles (see example in Figure 1) that is a weak potential for the priority rule. This is because, if a strategy profile $s$ is not an equilibrium, then $P$ increases along any of the shortest paths connecting it to an equilibrium, which in particular means that $s$ is not a local minimum point of $P$. ∎

## Sufficient Conditions for Weak Acyclicity

In a finite game $\Gamma$, a *never-best-reply* strategy for a player $i$ is a strategy $s_i$ that is not a best reply to any profile $s_{-i}$ of the other players' strategies. Since, as indicated, only pure strategies are considered here, every strategy that is strictly dominated by a *mixed* strategy is a never-best-reply strategy but not conversely. *Iterated elimination of never-best-reply strategies* means a finite sequence $\Gamma^0, \Gamma^1, \dots, \Gamma^m$ of games ($m \geq 0$) such that $\Gamma^0 = \Gamma$, each of the subsequent games is a subgame of the preceding game obtained by eliminating one or more never-best-reply strategies for one or more players, and there are no never-best-reply strategies in $\Gamma^m$. (It can be shown that the order of elimination does not matter, in the sense that the last subgame $\Gamma^m$ is unique. Obviously, the other subgames are also unique if in each step *all* eligible strategies are eliminated.) As the following lemma shows, the elimination process in a sense preserves the best-reply relation.

**Lemma 1** Consider iterated elimination of never-best-reply strategies in a finite game $\Gamma$, that is, a sequence of subgames $\Gamma^0$ ($= \Gamma$), $\Gamma^1, \dots, \Gamma^m$ as above. For $0 \leq l \leq m$, and any player $i$ and strategy profile $s$ in $\Gamma^l$, the player's strategy $s_i$ is a best reply to $s_{-i}$ in $\Gamma^l$ if and only if this is so in $\Gamma$. Moreover, for $0 \leq l \leq m$, the set of all equilibria in $\Gamma^l$ coincides with that in $\Gamma$.

*Proof.* For $0 \leq l \leq m$, player $i$ and strategy profile $s$ in $\Gamma^l$, if $s_i$ is *not* a best reply to $s_{-i}$ in $\Gamma^l$, then the same is obviously true in $\Gamma$. Conversely, if in $\Gamma$ player $i$'s strategy $s_i$ is not a best reply to $s_{-i}$, then consider any strategy $s_i'$ that is a best reply. Since the games $\Gamma^0, \Gamma^1, \dots, \Gamma^l$ all include every strategy in $s_{-i}$, they must also all include the best reply strategy $s_i'$, which is not eliminated. This implies that strategy $s_i$ is not a best reply also in the game $\Gamma^l$. The second part of the lemma follows from the first part and the fact that, for $0 < l \leq m$, every equilibrium in $\Gamma^{l-1}$ is present also in $\Gamma^l$, since each of the strategies in it is a best reply to the other strategies. ■

A game $\Gamma$ is *solvable* by iterated elimination of never-best-reply strategies if there is a sequence as above such that, in the last subgame $\Gamma^m$, all strategy profiles are equilibria. In this case, by Lemma 1, the set of all strategy profiles in $\Gamma^m$ coincides with the set of all equilibria in $\Gamma$. In other words, solvability means that the equilibria in $\Gamma$ are the only strategy profiles that survive iterated elimination of never-best-reply strategies.

**Theorem 2** (Kukushkin 2012, Apt and Simon 2015) If a finite game is solvable by iterated elimination of never-best-reply strategies, then it is weakly BI-acyclic.

*Proof.* Consider a game $\Gamma$ that is solvable by iterated elimination of never-best-reply strategies and a corresponding sequence of subgames $\Gamma^0, \Gamma^1, \dots, \Gamma^m$. Define the *height* of a strategy in $\Gamma$ as the largest index $l$ such that $\Gamma^l$ includes the strategy, and define the height of a strategy profile as the average height of its strategies. It suffices to establish the following.

CLAIM The function $P$ that maps strategy profiles to their height is a weak BI-potential for $\Gamma$. A strategy profile $s$ is an equilibrium if and only if $P(s) = m$.

To prove the claim, consider for a given strategy profile $s$ the minimum of its strategies' heights. Clearly, this minimum $k$ is equal to $m$ if and only if $s$ is a strategy profile also in $\Gamma^m$, which by the solvability assumption means that it is an equilibrium. If $k < m$, then all the strategies in $s$ are in $\Gamma^k$ but at least one of them, $s_i$, is not in $\Gamma^{k+1}$. Necessarily, for the corresponding player $i$, it holds (both in $\Gamma^k$ and in $\Gamma$; see Lemma 1) that strategy $s_i$ is not a best reply to $s_{-i}$. Let $s_i'$ be a strategy in $\Gamma^k$ that is a best reply. Its height is necessarily greater than $k$, and therefore the strategy profile $(s_i', s_{-i})$ satisfies $P(s_i', s_{-i}) > P(s)$. Thus, $s$ is not a local minimum point of $P$. The conclusion completes the proof of the claim and therefore also that of the theorem. ■

A different sufficient condition for weak acyclicity is presented by the following theorem. For an example, see Figure 1.

**Theorem 3** (Fabrikant et al. 2013) If a finite game has the property that every subgame has a unique equilibrium, then it is weakly BI-acyclic.

Interestingly, the weaker property that every subgame has *at least* one equilibrium is not sufficient even for weak I-acyclicity (Takahashi and Yamamori 2002).

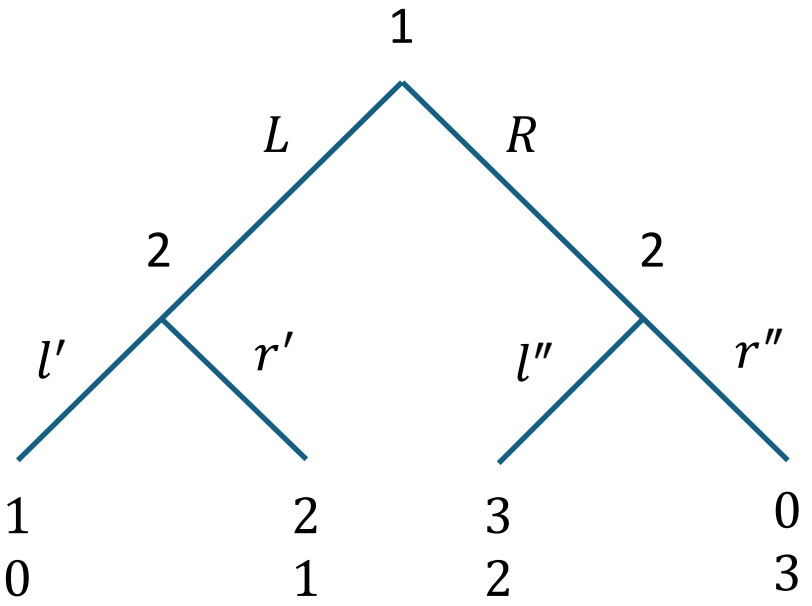

**FIGURE 2 An extensive-form game with perfect information that is not BI-acyclic. The best-improvement graph of the game has the following closed walk: $(L, r'l'')$, $(R, r'l'')$, $(R, l'r'')$, $(L, l'r'')$ and back to $(L, r'l'')$.**

## Extensive-Form Games

The difference between acyclicity and weak acyclicity is illustrated by the example of finite extensive-form games with perfect information. In general, these games are not even BI-acyclic, as the simple game in Figure 2 demonstrates. However, this particular counterexample is clearly driven by player 2's nonbeneficial change of action at his unreached decision nod (simultaneously with the beneficial change at the reached nod). Indeed, in the subgame whose root is the unreached nod, the change of strategy is harmful. In other words, the agent (Selten 1975) residing at that node switches to an action that would not be optimal if the node were actually reached. The significance of this observation lies in the fact that the agent normal form of every finite extensive-form game with perfect information *is* acyclic. This fact can be stated also as follows.

**Theorem 4** (Kukushkin 1999) Every finite extensive-form game with perfect information in which each player has only one decision node is I-acyclic.

Theorem 4 is a special case of a more general result, which applies regardless of the numbers of the players' decision nods. Namely, if a walk in the improvement graph does not involve changes of actions at unreached decision nods as in Figure 2, then it cannot be closed (Kukushkin 2002). Put differently, the priority rule that forbids changes of strategies involving such changes of actions is acyclic. An alternative (similar, but a trifle stronger) acyclic priority rule can be defined as follows. For a player $i$ and a strategy profile $s$, call a unilateral change of strategy from $s_i$ to another strategy $s_i'$ *parsimonious* if $u_i(s_i', s_{-i}) > u_i(s_i'', s_{-i})$ for every strategy $s_i''$ that differs from both $s_i$ and $s_i'$ but is a combination of them, in the sense that the action it prescribes at each of player $i$'s decision nodes is also prescribed by one (or both) of these strategies. Clearly, any non-parsimonious change of strategy can be replaced by a parsimonious one without decreasing the player's payoff as a result. Therefore, the subgraphs of the improvement- and best-improvement graphs obtained by considering only parsimonious changes of strategies are priority rules.

**Lemma 2** In a finite extensive-form game with perfect information $G$, the I-priority rule and the BI-priority rule that allow only parsimonious changes of strategies are acyclic.

*Proof.* It has to be shown that every walk $s^0, s^1, \ldots, s^m$ in the I-priority rule (or, as a special case, the BI-priority rule) under consideration is not closed. For a subgame $\hat{G}$ of $G$ and a strategy profile $s$, denote by $\hat{s}$ the strategy profile in $\hat{G}$ obtained by restricting each player's strategy to the decision nodes in that subgame. In particular, $\hat{s}^0, \hat{s}^1, \ldots, \hat{s}^m$ are the strategy profiles corresponding to the walk $s^0, s^1, \ldots, s^m$. Now, choose $\hat{G}$ in such a way that there is exactly one player $i$ whose strategies $\hat{s}_i^0, \hat{s}_i^1, \ldots, \hat{s}_i^m$ are not all equal. For any $0 \le l < m$ such that $\hat{s}_i^l \neq \hat{s}_i^{l+1}$, consider the strategy $s_i^{l+1/2}$ ($\neq s_i^{l+1}$) in $G$ that coincides with $s_i^l$ inside $\hat{G}$ and with $s_i^{l+1}$ outside it. Since the chance from $s_i^l$ to $s_i^{l+1}$ increases $i$'s payoff and is parsimonious, $u_i(s_i^{l+1/2}, s_{-i}^l) < u_i(s^{l+1})$. The inequality implies that, in the subgame $\hat{G}$ , strategy $\hat{s}_i^{l+1}$ yields a higher payoff to player $i$ than $\hat{s}_i^l$ does, when the other players' strategies are $\hat{s}_{-i}^l$. Since by assumption the latter strategies do not change (that is, $\hat{s}_{-i}^0 = \hat{s}_{-i}^1 = \cdots = \hat{s}_{-i}^m$), the conclusion implies that the walk cannot be closed. ■

A change of strategy that changes the action only at a single decision node is (vacuously) parsimonious. Therefore, Theorem 4 is an immediate corollary of Lemma 2. However, as the lemma guarantees the existence of an acyclic priority rule also in the general case, its usefulness is not limited to the special games considered in that theorem. It also provides an alternative proof for the following result.

**Theorem 5** (Kukushkin 2002) Every finite extensive-form game with perfect information is weakly BI-acyclic.

By Theorems 4 and 5, acyclicity or weak acyclicity of a finite game $\Gamma$ is a necessary condition for the existence of *some* perfect-information extensive-form game $G$ whose agent normal form or normal form, respectively, is $\Gamma$. If $\Gamma$ is the agent normal form of $G$, then it clearly has the stronger property that every subgame is acyclic (see footnote 1). However, if $\Gamma$ is the normal form, then it may have a subgame that is not weakly acyclic. For example, in the normal form of the game in Figure 2, the subgame where player 2 is only allowed to use strategies $r'l''$ and $l'r''$ does not even have an equilibrium.